\documentclass[epsf,preprint,aps]{revtex4}
\usepackage{latexsym}
\usepackage{amsfonts}
\usepackage{graphicx}
\usepackage{epsfig}
\usepackage{amsmath, amsthm, amssymb}

\begin{document}
\title{Dynamics of 2D Monolayer Confined Water in Hydrophobic and
  Charged Environments}

\author{Pradeep Kumar}

\affiliation{ Center for Studies in Physics and Biology, The
  Rockefeller University, 1230 York Avenue, New York, NY 10021 USA
  \\ }

\begin{abstract}

Using molecular dynamics simulations we study the dynamics of a
water-like TIP5P model of water in hydrophilic and hydrophobic
confinement. We find that in case of extreme nanocofinement such that
there is only one molecular layer of water between the confinement
surface, the dynamics of water remains Arrhenius with a very high
activation energy up to high temperatures. In case of polar
(hydrophilic) confinement, The intermediate time scale dynamics of
water is drastically modified presumably due to the transient coupling
of dipoles with the effective electric field due to the surface
charges. Specifically, we find that in the presence of the polar
surfaces, the dynamics of monolayer water shows anomalous region --
namely the lateral mean square displacement displays a distinct
superdiffusive intermediate time scale behavior in addition to
ballistic and diffusive regimes. We explain these finding by proposing
a simple model. Furthermore, we find that confinement and the surface
polarity changes the vibrational density of states specifically we see
the enhancement of the low frequency collective modes in confinement
compared to bulk water. Finally, we find that the length scale of
translational-orientational coupling increases with the strength of
the polarity of the surface.

\end{abstract}

\maketitle

\noindent
\section{Introduction}
\bigskip

Water is the most ubiquitous liquid and hence plays a very important
role in different phenomena. The list of such phenomena ranges across
as varied disciplines as physics, chemistry, biology, nanofluidics,
geology and atmospheric
sciences~\cite{leshouches,Ball2008,Kumar06PRL,Buldyrev07,Michaelides2007,Carrasco2009,Rasaiah2008}. Water
also falls into a class of complex liquids which are known to have
anomalous behavior compared to simple
liquids~\cite{FrankBook,Debenedetti2003}. The anomalous expansion upon
decreasing temperature below $4^{o}$~C and increase of specific heat
upon decreasing temperature in the supercooled state are some of the
examples of anomalies of water. Indeed, the list of the anomalous
behaviors is incomplete and seems to be ever
growing~\cite{Stanley07,chaplin}.

Studies of water confined in carbon nanotubes and between hydrophobic
surfaces suggest that the confining surfaces may induce layering in
liquid water in extreme nano
confinements~\cite{Zangi2004,Koga2003}. Moreover, first order layering
transitions are also observed~\cite{Zangi2004,Koga2000} in hydrophobic
confinements. Water can also form wide variety of crystalline phases
such as monolayer and bilayer
structures~\cite{Koga2000,Koga2001,Koga2003,Zangi2004,zangi1,zangi2,Kumar2005}
in nanoconfinements. Apart from distinct structural changes, water
also exhibits dynamic changes depending on the nature of the confining
substrates. It has been found the dynamics becomes slow near
hydrophilic surfaces compared to hydrophobic surfaces, presumably due
to the binding of polar oxygens and hydrogens to the polar groups on
the hydrophilic substrates. Moreover, the dynamics in confinement may
depend on the surface morphology. Specifically, it was found that a
liquid in smooth confinements diffuses faster than when the confining
surfaces are rough~\cite{kb1}. Recent studies
~\cite{Kumar2005,Gallo07,Brovchenko07,Han2008a,truskett} suggest that
both thermodynamic and dynamic anomalies of water are strongly
affected when water is confined in small nanopores. It is found that
when water is confined between hydrophobic surfaces both thermodynamic
and dynamic anomalies shift to low temperatures and low
pressures~\cite{Kumar2005,Gallo07}. Although the characteristics of
hydrogen bond dynamics of water is bulk-like in hydrophobic
confinement, the average life-time decreases~\cite{Han2008b} leading
to a shift in the regions of anomalous dynamics and thermodynamics in
the P-T
plane~\cite{Kumar2005,Gallo07,Brovchenko07,truskett}. Moreover, the
long time relaxation of hydrogen bonds depend on the effective system
dimensionality~\cite{Han2008b}.

One of the many important questions that can be asked from the
biological science perspective is -- what role, if at all, water plays
in biology ?  The answer to such questions is probably not very
apparent and requires a systematic effort to understand physical
behavior of water under various conditions such as -- water in and
around ion channels, protein-hydration water, water in different
solutions etc. Although there has been a large array of works
answering some of the
questions~\cite{chandler1,protein,Hummer00,Chen05,Liu04,xuPNAS,Buldyrev07,Hummer03,Kumar06PRL,Brovchenko06,Hanasaki06,Raviv2001,Berne04,nicolas1,nicolas2},
the behavior of water is far from fully understood.

In this paper, we study the dynamics of a monolayer water confined
between both hydrophobic and polar surfaces. In section II, we
describe the system and method and discuss the results of positive and
negative surface polarity on the dynamics of water in section III. In
section IV, we present the results of the orientational dynamics and
translational-orientational coupling and finally we conclude with a
discussion and summary in section V.

\bigskip
\section{System and Method}
\bigskip

\begin{figure}
\begin{center}
\includegraphics[width=14cm]{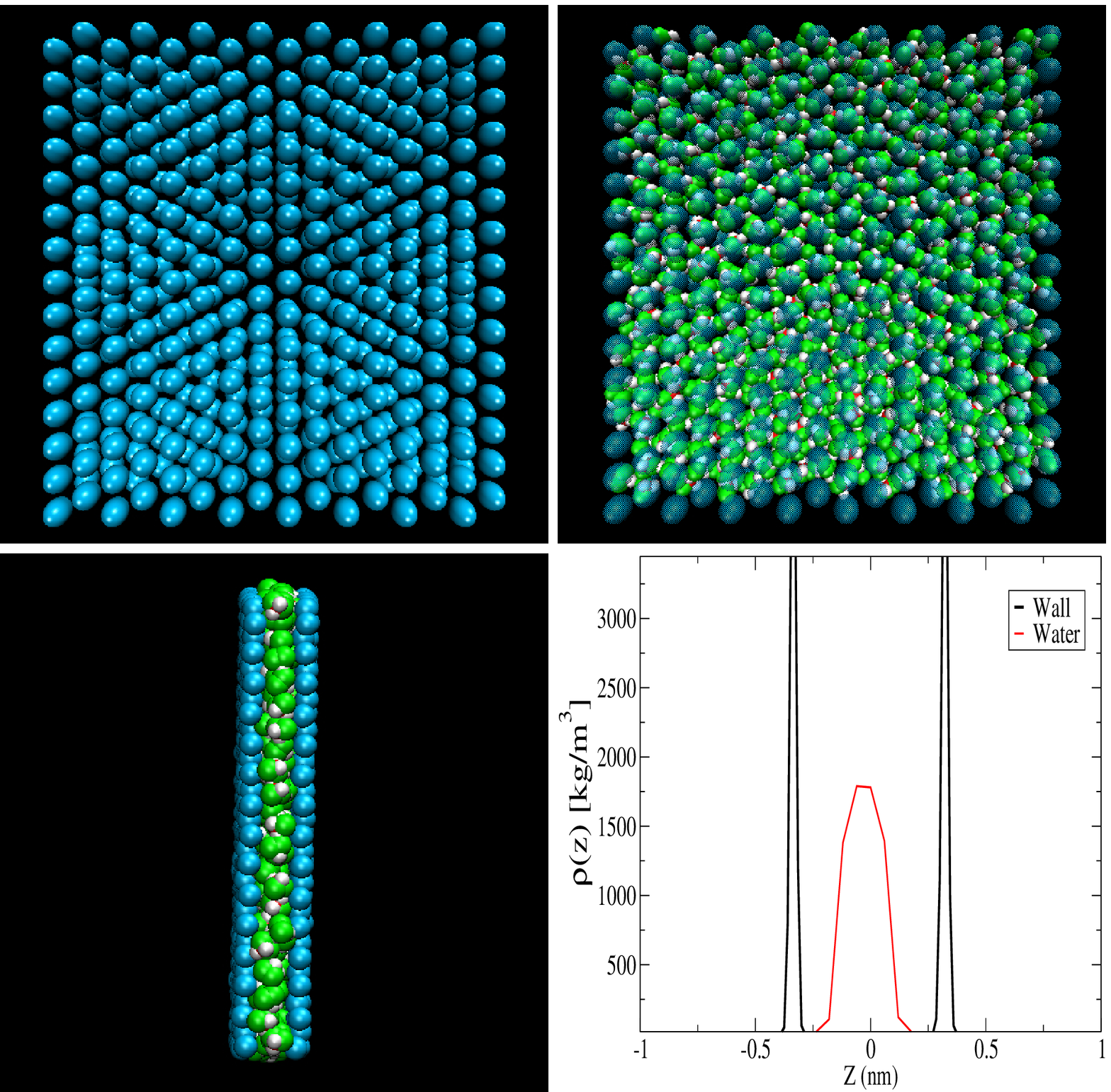}
\end{center}
\caption{(Color online) Schematics of monolayer water confined between two
  surfaces. (a) The top view of the confining surfaces. (b) The top
  view of the confining surface with the water molecules. (c) The
  lateral view of the confined system. (d) Density profile $\rho(z)$
  of the surface atoms and the water molecules along the confinement
  direction. The effective confinement width due to excluded volume
  interaction between the water molecules and the surface molecules is
  $~0.320$nm, which is about the diameter of a water molecule.}
\label{fig:schematic}
\end{figure}

We perform the molecular dynamics simulation of TIP5P (transferable
interaction potential five
points)~\cite{JorgensenXX,jorgensen2,YamadaXX,Paschek2005} water-like
molecules confined between two structured surfaces separated by
$0.6$~nm. The separation is chosen such that there is effectively one
layer of water between the surfaces (see
Fig.~\ref{fig:schematic}). The atoms on the surfaces are arranged in a
hexagonal packing with $\sigma_S=0.339$~nm and do not interact with
each other.  The positions of the surface atoms were constrained to
their respective mean positions by a harmonic potential with spring
constant $100$~kJ/mol/nm$^2$. In the case of hydrophobic surface, the
interaction between the surface atoms and the oxygen of the water
molecule is modeled using 6-12 LJ potential U(r)
\begin{equation}
U(r) = 4\epsilon[(\frac{\sigma}{r_{SO}})^{12} -
  \frac{\sigma}{r_{SO}})^{6}]
\end{equation}
where $r_{SO}$ is the distance between the surface atom and oxygen of
water molecule, and $\sigma = (\sigma_{S}+\sigma_{OW})/2$ and
$\epsilon = \sqrt{\epsilon_{S}\epsilon_{OW}}$. $\sigma_{OW}$ and
$\epsilon_{OW}$ are the parameters of the LJ interaction between
oxygens of water molecules~\cite{JorgensenXX,jorgensen2}. The
hydrophilicity of the surface is achieved by adding point charge
$Qe$~Coulomb to each surface atom and in that case long-range
Coulombic interactions between surface atoms and water molecules are
added in addition to the LJ interaction. We performed simulations of
hydrophobic and hydrophilic confined water in NVT ensemble with the
effective density of water $\rho=1.00$~g/cm$^3$. The equations of
motion are integrated with a time step of $0.001$~ps and Berendsen's
thermostat is used to attain constant temperature.

\section{Translational Dynamics and Vibrational Density of States}
\bigskip

The translational dynamics in the confined systems can be described by
lateral mean-square displacement (MSD) $\langle r_{||}^2(t)\rangle$,
parallel to surface in the periodic directions. In 2-dimension, the
diffusion constant $D_{||}$ can be calculated by using a modified
Einstein relation between the long time behavior of the $\langle
r_{||}^2(t)\rangle$ and $D_{||}$~\cite{hansen}
\begin{equation}
D_{||} = \lim_{t\rightarrow\infty} \frac{\langle  r_{||}^2(t)\rangle}{4t}
\end{equation}

\begin{figure}
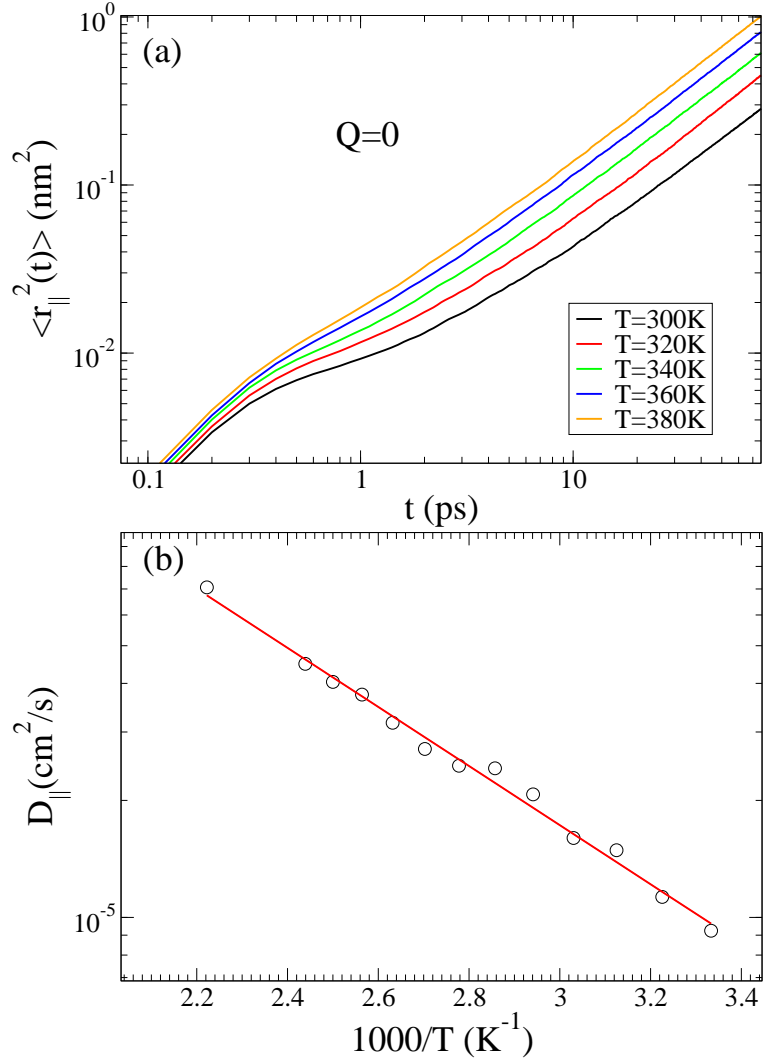

\begin{center}
\includegraphics[width=10cm]{fig2a.eps}
\includegraphics[width=10cm]{fig2b.eps}
\end{center}
\caption{(Color online) (a) $\langle r_{||}^2(t)\rangle$ as a function of $t$ for
different temperatures when there are no charges on the surface. (b) A
linear dependence of $D_{||}$ on $1/T$ on Arrhenius plot shows that
the dynamics remains Arrhenius even at high temperatures with an
activation energy $E_{A}= 14.12$~kJ/mol.}
\label{fig:msdq0}
\end{figure}

We first study the temperature dependence of $\langle
r_{||}^2(t)\rangle$ and $D_{||}$ in the case when there are no charges
on the surface. In Fig.~\ref{fig:msdq0} (a), we show the MSD $\langle
r_{||}^2(t)\rangle$ for different temperatures. The MSD shows typical
time scales --namely a ballistic regime at very small time scales, a
cage regime at intermediate time scales, and diffusive regime at long
times. In Fig.~\ref{fig:msdq0}, we show the Arrhenius plot of $D_{||}$
extracted from the long time linear regime of the MSD. We find that
the dynamics remains Arrhenius for all the temperature range studied
and can be fit with $D_{||}\equiv D_{||}(0)e^{-E_A/k_BT}$, where
$E_A=14.12$~kJ/mol and $k_B$ are the activation energy and the
Boltzmann constant respectively and $D_{||}(0)$ is a fitting
parameter. An Arrhenius behavior of $D_{||}$ suggests the the water
remains very ordered upto high temperatures in extreme hydrophobic
confinements. Moreover, the activation energy is comparable to the low
density liquid (where the local structure resembles the IceIh) of bulk
water phase.

\begin{figure}
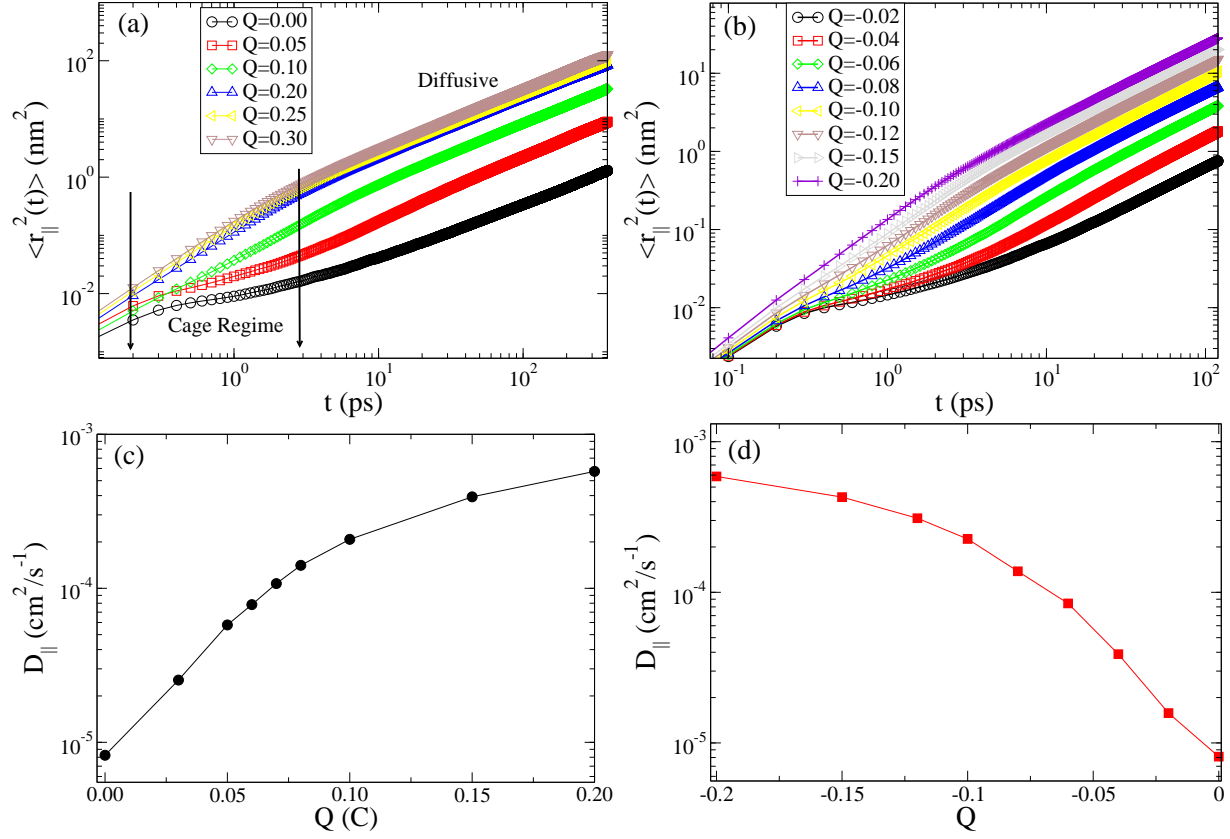

\begin{center}
\includegraphics[width=8cm]{fig3a.eps}
\includegraphics[width=8cm]{fig3b.eps}
\includegraphics[width=8cm]{fig3c.eps}
\includegraphics[width=8cm]{fig3d.eps}
\end{center}
\caption{(Color online) (a) Lateral mean square displacement $\langle
  r_{||}^2(t) \rangle$ as a function of $t$ for different values of
  $Q$. Increasing polarity changes the intermediate time scale
  dynamics -- namely the slope of the cage region increases as the
  polarity increases. For large values of $Q$, the cage regime becomes
  superdiffusive. (b) $\langle r_{||}^2(t)\rangle$ for the case of
  surface with negative polarity.  The behavior in the case of
  negative polarity is similar to the case of positive polarity (see
  Fig.~\ref{fig:msd} (a)) (c) Lateral diffusion constant $D_{||}$ of
  water molecules increases with increasing $Q$ and seems to
  asymptotically saturate to a constant value for large $Q$. (d)
  Analog of Fig.~\ref{fig:msd}(c) for negative surface polarity.}
\label{fig:msd}
\end{figure}

We next study the translational dynamics in the case when surface
atoms have charges on them. In Figure~\ref{fig:msd}(a) and (b), we
show the MSD displacement at $T=300$~K as a function $t$ for different
strength of surface polarity. For the hydrophobic case when $Q=0$, we
find that the dynamics of water in confinement is much slower with a
lateral diffusion constant
$D_{||}\approx8.0\rm{x}10^{-6}\rm{cm}^2/\rm{s}$. Note that the
diffusion constant of TIP5P bulk water at the same density and
temperature is $\approx
~2.5\rm{x}10^{-5}\rm{cm}^2/\rm{s}$~\cite{YamadaXX}. When the polarity
of the surface increases, there are namely two changes in the behavior
of the MSD -- (i) the change in the behavior of cage dynamics or the
intermediate time scale dynamics and (ii) increase of the long time
MSD and hence the diffusion constant $D_{||}$. In
Fig.~\ref{fig:msd}(c) and (d), we show $D_{||}$ for positive and
negative polarity of the surfaces respectively. As we would expect the
dependence of $D_{||}$ on strength of polarity in both the cases are
similar. 

Moreover, we find that dynamics in polar confinements is different
from dynamics at high $T$ in apolar confinement. A comparison of MSD
at high $T$ and $Q=0$ with MSD for $T=300$ and $Q=0.06$ suggests that
while at high T and $Q=0$ the MSD changes directly from ballistic to
diffusive behavior, the MSD shows a distinct difference at the
intermediate time scales in case of polar confining surfaces.

\begin{figure}
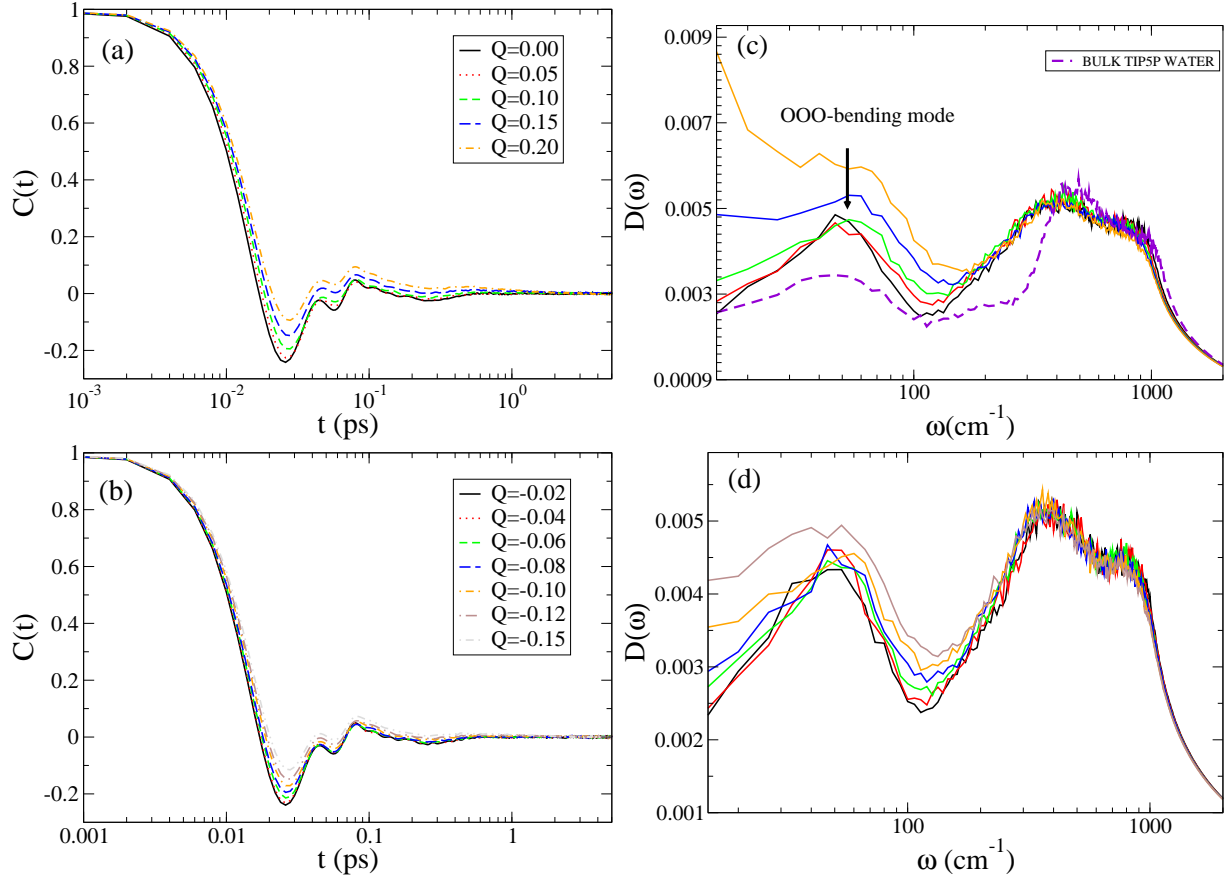

\begin{center}
\includegraphics[width=8cm]{fig5a.eps}
\includegraphics[width=8cm]{fig5b.eps}
\includegraphics[width=8cm]{fig5c.eps}
\includegraphics[width=8cm]{fig5d.eps}
\end{center}
\caption{(Color online) (a) Velocity autocorrelation function $C(t)$ as a function
  of time for different values of the polarity of the surface. The
  anticorrelation in $C(t)$ begins to disappear as the polarity of the
  surface increases. (b) Analogue of Fig.~\ref{fig:vdos}(a) for
  negative polarity. (c) Vibrational density of states as calculated
  from the fourier transform of $C(t)$. A comparison with bulk water
  $D(\omega)$ shows that the hydrogen bond stretch band ($\approx
  210~\rm{cm}^{-1}$) is absent in the confinement. Note the merging of
  diffusive modes into low frequency collective modes when $Q$
  increases. (d) Analog of Fig.~\ref{fig:vdos}(c) for negative
  polarity.}
\label{fig:vdos}
\end{figure}

The change in the behavior of the intermediate time scale dynamics can
be understood by the fact that although the time average of external
electric field at a point inside the confined region is zero, a
transient coupling of the fluctuating surface electric field
$\vec{E}(\vec(x),t)$ with the dipole $\vec{\mu}(\vec{x},t)$ of a
water molecule at position $\vec{X}$ will impart a torque
$\vec{\mu}(\vec{x},t)X\vec{E}(\vec{x},t)$. The electric field
$\vec{E}(\vec{x},t)$ due to one of the surfaces at a position
$\vec{x}$ is given by
\begin{equation}
\vec{E}(\vec{x},t) = \frac{1}{4\pi\epsilon}
\sum_{i=1}^{N_q}Q_i\frac{\vec{x}-\vec{x}_i}{|\vec{x}-\vec{x}_i|^3}
\label{eq:fullE}
\end{equation}
where the sum is taken over $N_q$ charges on the surface and
$\vec{x}_i$ are the position vectors of the point charges on the
surface. The dynamics of $x_i$ follows the dynamics of a Brownian
harmonic oscillator and can be written as:
\begin{equation}
\frac{dv_i}{dt} = -m\omega^2x_i - \zeta v_i + F(t),
\label{eq:harmonic}
\end{equation}
where $F(t)$ is a random force and $\zeta$ is the frictional
attenuation due to water. Note that in our system the surface atoms
do not interact with each other. Although the set of two equations
describes the time dependence of surface electric field, it is almost
intractable. To simplify this, we assume that the electric field
inside the confined region is due to a flat surface with a fluctuating
surface charge density $\sigma(t)$ and hence $\vec{E}(t)$ now will be
perpendicular to the surface and can approximately be written as:

\begin{equation}
\vec{E}(t) = \frac{\sigma(t)}{2\epsilon}\hat{z}
\end{equation}
where the surface charge density $\sigma(t)$ is 
\begin{equation}
\sigma(t) = \frac{N_qQ}{A(t)}
\end{equation}
where $A(t)$ is the surface area at a given time $t$. Note that the
expression for electric field is approximate but we assume that the
extent of the surface area considered for $\sigma(t)$ is very large
compared to the distance of the point at which we are interested to
find the electric field at. Let's further assume that
$A(t)=\bar{A}(1-\frac{\delta A(t)}{\bar{A}})$, we can write the
instantaneous fluctuations $\delta \sigma(t)$ in $\sigma$ as:
\begin{equation}
\delta \sigma(t) = \frac{N_qQ}{\bar{A}^2}\delta A(t),
\label{eq:dsigma}
\end{equation}
where $\bar{A}$ is the average area of consideration. Using
Eq.~\ref{eq:dsigma} we can the time autocorrelation of fluctuations in
$\sigma(t)$ as:
\begin{equation}
\langle \delta \sigma(t) \delta \sigma(0) \rangle =
\frac{N_q^2Q^2}{\bar{A}^4} \langle \delta A(t) \delta A(0) \rangle
\end{equation}
Hence the time autocorrelation in the instantaneous fluctuations
$\delta E$ of $E$ can be written as:
\begin{equation}
\langle \delta E(t) \delta E(0) \rangle = \frac{N_q^2Q^2}{4\epsilon^2
  \bar{A}^4}\langle \delta A(t) \delta A(0) \rangle
\end{equation}
Now assuming that $A=Lx.Ly=(x_{n}-x_{1})(y_{n}-y_{1})=xy$, where
$x_1,y_1$ and $x_n,y_n$ are the coordinates of the $1^{st}$ and
$n^{th}$ point charge along the boundary of the area (note that for
simplicity we only cosider the in-plane motion of the point
charges). The dynamics of the new coordinates $x$~ and $y$ are given
by Eq.~\ref{eq:harmonic}. $\langle \delta E(t)\delta E(0) \rangle$ can
now be written as:
\begin{equation}
\langle \delta E(t) \delta E(0) \rangle = \frac{N_q^2Q^2}{2\epsilon^2\bar{A}^3} \langle
\delta x(t) \delta x(0) \rangle = \bar{E}^2\frac{2}{\bar{A}}  \langle \delta x(t) \delta x(0) \rangle,
\end{equation}
where $\delta x = \bar{x}-\delta x(t)$~is the instantaneous
fluctuation in $x(t)$ with $\langle \delta x(t) \rangle = 0. $For
simplification in the above equation, we have assumed
$\bar{x}=\bar{y}$. Using Eq.~\ref{eq:harmonic} for coordinate $\delta
x$, the time autocorrelation of $\delta E$ can be written as:
\begin{equation}
\langle \delta E(t) \delta E(0)\rangle = \bar{E}^2\frac{2}{\bar{A}}\frac{k_BT}{m\omega_{0}^2}e^{-\frac{\zeta}{2m}t}(cos\omega_1t+\frac{\gamma}{2m\omega_1}sin\omega_1t), 
\end{equation}
where $\omega_{0}=\sqrt{\frac{k}{m}}$ is the characteristic frequency
of point charges with spring constant $k$ and mass $m$, and
$\omega_{1} = \sqrt{\omega_{0}^2-\frac{\zeta^2}{4m^2}}$. We can write
the time autocorrelation of instantaneous fluctuation $\delta
E_{T}(t)$ in the net electric field can be written as:
\begin{equation}
\langle \delta E_{T}(t) \delta E_{T}(0)\rangle = \bar{E}^2\frac{4}{\bar{A}}\frac{k_BT}{m\omega_{0}^2}e^{-\frac{\zeta}{2m}t}(cos\omega_1t+\frac{\gamma}{2m\omega_1}sin\omega_1t), 
\label{eq:final}
\end{equation}
From Eq.~\ref{eq:final}, we can see that the fast relaxation of
autocorrelation of fluctuations in total electric field at a point in
the confined region occurs with relaxation time $2m/\zeta$. One would
expect that over this time scale the net electric field would not
completely vanish in the confined region due to both surfaces and
hence there would be a transient coupling of water dipoles with the
net electric field. If this energy of dipole coupling is comparable to
small frequency collective modes, such transient coupling of dipoles
with the surface electric field will lead to a change in the
intermediate time scale dynamics. Assuming a surface with uniform
charge distribution, a simple estimate shows that $Q = 0.10$~ would be
equivalent to an effective electric field~$bar{E}\approx
10^8$~V/m. Assuming the length of the considered area to be twice the
cut-off length ($\approx$ 2nm), the standard deviation of the net
electric field is $\approx 0.22\bar{E}$ at $T=300$~K. In this case,
the field-dipole coupling would be large enough to change the low
frequency collective vibrations. To verify our picture we calculate
the vibrational density of state $D(\omega)$ from the Fourier
transform of the velocity autocorrelation function $C(t)$ defined as
\begin{equation}
D(\omega) \equiv \int_{0}^{\infty}C(t) e^{i\omega t} dt,
\label{eq:vdos}
\end{equation}
where $\omega$ is the frequency and $C(t)$ is defined as
\begin{equation}
C(t) \equiv
\frac{\langle \vec{v}(t).\vec{v}(0)\rangle}{\langle \vec{v}(0).\vec{v}(0)\rangle}
=  \frac{1}{Nv(0)^2}\sum_{i=1}^{N}
\langle  \vec{v}_i(t).\vec{v}_i(0)\rangle.
\end{equation}
where $v_i(t)$ denotes the velocity of the $\rm{i}^{\rm{th}}$-water
molecule at time $t$ and angular brackets denote the ensemble
average. In Fig.~\ref{fig:vdos}(a) and (b), we show $C(t)$ for
different values of positive and negative polarity of the surface
respectively. We find that as the magnitude of the polarity increases,
the anticorrelation in $C(t)$ at $t\approx 0.03$~ps decreases. In
Fig.~\ref{fig:vdos} (c) and (d), we show the vibrational density of
states obtained using equation Eq.~\ref{eq:vdos}. A comparison of the
bulk TIP5P and confined vibrational density of state shows that the
low frequency collective modes are enhanced in confinement.
Furthermore, the hydrogen bond stretch mode ($\approx 210$ cm$^{-1}$)
is absent (or not distinguishable) in the extreme confinement case
(see Fig.~\ref{fig:vdos}). We find that when magnitude of the polarity
increases, the peak corresponding to the OOO bending mode ($\approx
50\rm{cm^{-1}}$) gets broader and merges into the diffusive modes
suggesting that highly charged surfaces can modify low frequency
collective modes and hence may lead to destruction of the local order.

\section{Orientational Dynamics}
\bigskip

We next study the orientational dynamics of water molecules since we
expect that the transient coupling of fluctuating surface electric
field to dipoles of water would provide exteral torque which would
tend to align the water dipoles. To study the orientational dynamics
of water, we calculated the self-dipole orientational correlation
function $C_{d}(t)$~\cite{hansen} defined as
\begin{equation}
C_{d}(t) = \langle 
\frac{\vec{\mu}(t).\vec{\mu}(0)}{\vec{\mu}(0).\vec{\mu}(0)} \rangle =
\frac{1}{N} \sum \langle  \cos \theta_i(t)\rangle
\end{equation}
where $\vec{\mu}_i(t)$ is the dipole of $\rm{i}^{\rm{th}}$ water
molecule at time $t$, $\cos\theta_i (t)$ is the angle between the
dipole vectors of $\rm{i}^{\rm{th}}$-molecule at time $t$ and time
$0$.

\begin{figure}
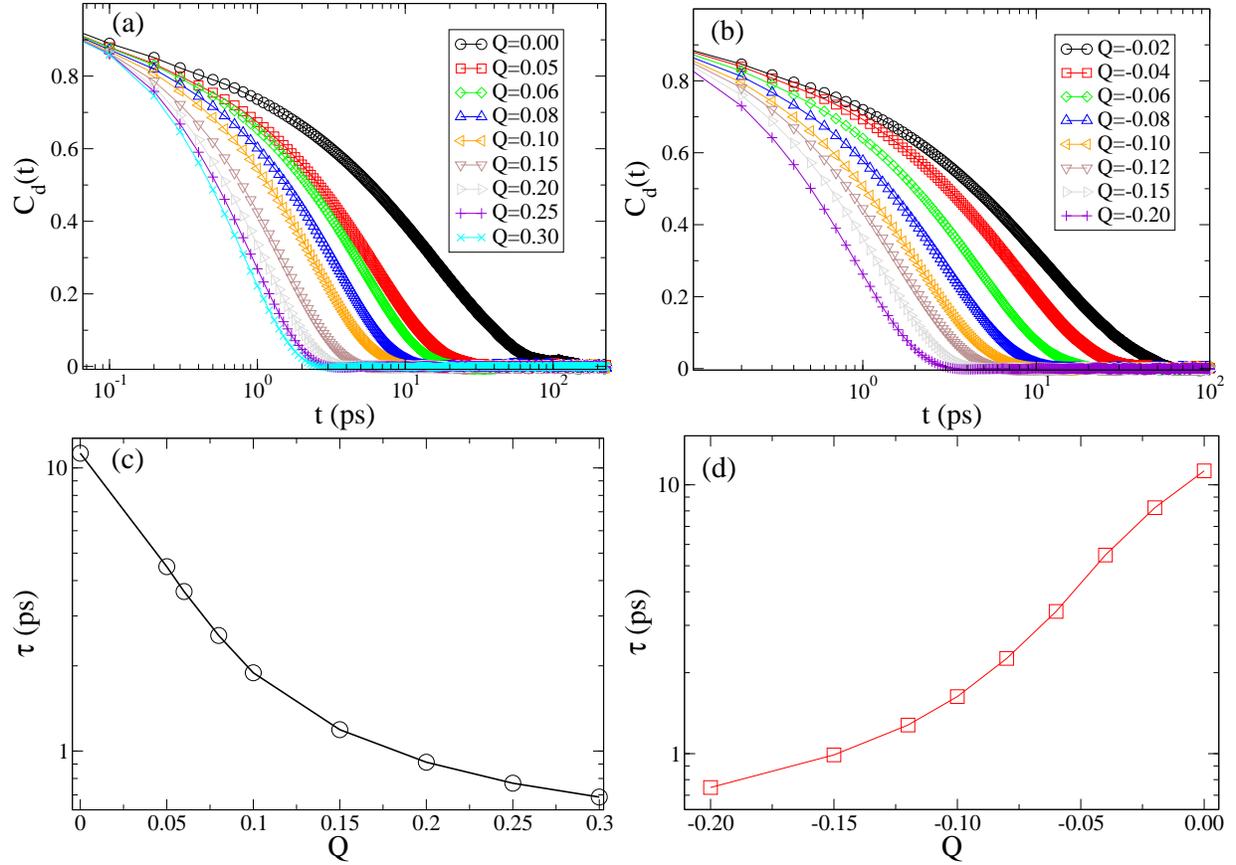

\begin{center}
\includegraphics[width=8cm]{fig6a.eps}
\includegraphics[width=8cm]{fig6b.eps}
\includegraphics[width=8cm]{fig6c.eps}
\includegraphics[width=8cm]{fig6d.eps}
\end{center}
\caption{(Color online) (a) Dipole-dipole orientational correlation function $C(t)$
  as a function of $t$ for different values of the polarity. C(t)
  decays faster as the surface polarity is increased. (b) Analog of
  Fig.~\ref{fig:ddcorr}(a) for negative surface polarity. (c)
  Orientational correlation time $\tau$ as a function of $Q$ suggests
  that the orientational correlation time increases monotonically. (d)
  Analog of Fig.~\ref{fig:ddcorr}(c) for negative surface polarity.}
\label{fig:ddcorr}
\end{figure}

\noindent In Fig.~\ref{fig:ddcorr} (a) and (b), we show $C_{d}(t)$ for
different positive and negative values of $Q$. We find that, for small
magnitude of $Q$, the orientational correlation function $C_{d}(t)$
decays to zero slowly. To quantify the characteristic time of
orientational correlations, we define correlation time $\tau$ as the
time at which $C_{d}(t)$ decays by a factor of $e$. In
Fig.~\ref{fig:ddcorr} (c) and (d), we show $\tau$ for the positive and
negative surface polarity respectively. We find that, as the magnitude
of $Q$ increases, $\tau$ decreases and seems to saturate to a constant
value for larger magnitudes of $Q$.

\medskip

\begin{figure}
\begin{center}
\includegraphics[width=12cm]{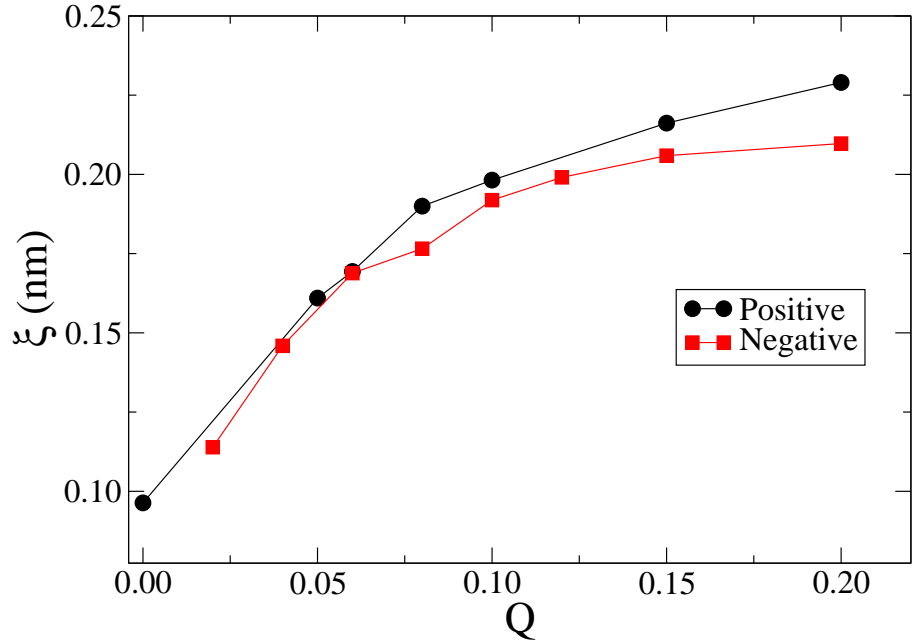}
\end{center}
\caption{Translation-orientational coupling length scale $\xi$ as a
  function of atomic charge on the surface. The circles and squares
  are for positive and negative polarity respectively. $\xi$ increases
  as the the magniture of the charges increases for both polarities of
  the surface.}
\label{fig:xi}
\end{figure}

\noindent A significant coupling of translational and orientational
motion is found in bulk water~\cite{Chen1997,Mazza2006} and other
confined systems~\cite{Joseph2008}. Since we can expect that the
presence of the charges of the confining surfaces strongly influence
the dipole fluctuations of the confined water, we would also expect
that changing the polarity may lead to a change in the effective
length scale of the coupling of the translational and orientational
dynamics. The characterize the the effective length scale of the
coupling between translational and orientational motion, we define
$\xi$
\begin{equation}
\xi = \sqrt{\rm{D}_{||}*\tau}
\end{equation}
We interpret $\xi$ as the effective length scale over which
translational and orientational motions are coupled. Figure
~\ref{fig:xi} shows $\xi$ for different values of $Q$ for both
positive and negative $Q$. The value $\xi$ increases as the $Q$ is
increased and seems to aysmptotically reach a constant
value. Furthermore, we find that the value of $\xi$ is similar for
both positive and negative surface polarities.
\bigskip

\section{Summary and Discussions}

In summary, using molecular dynamics simulations of TIP5P model of
water, we have investigated the dynamics of a monolayer water system
confined between hydrophobic and hydrophilic surfaces of different
polarity. We find that dynamics of monolayer water in hydrophobic
confinement remains Arrhenius upto very large temperatures with a very
high activation energy $E_A=14.12$~kJ/mol. Further, we find that the
transient coupling of the water dipoles with the electric field of the
surface gives rise to distinct cage dynamics. Specifically at high
polarity the slope of the MSD as a function of $t$ at the intermediate
time scales becomes $>1.0$ on a log-log scale, giving rise to three
distinct dynamic regions --ballistic at very small time scales,
superdiffusive at intermediate times and a long time diffusive
behavior. Note that although the effect of transient coupling of water
dipoles to the external field is effective increase of diffusion of
the system, the dynamic changes are very different in the two
cases. While at high temperatures MSD would only show two distinct
regimes, water in polar confinements has a different intermediate time
scale dynamics and three distinct regimes.  We explain these finding
by proposing a simple model of electric field fluctuations in the
confined region. We further find that the extreme confinement enhances
the low frequency collective modes. Moreover, we find that the
presence of a highly polar surface disrupts the low frequency
collective vibrational modes and hence leads to less ordered system,
suggesting that while extreme hydrophobic confinement increases the
order, the presence of a highly polar surface makes the water molecule
relatively more disordered in confinement. Finally, we calculated the
effective length scale of coupling of translational and orientational
motions and find that the coupling length scale increases as the
polarity of the surface increases.

\newpage

\newpage

\newpage

\newpage

\begin{figure}
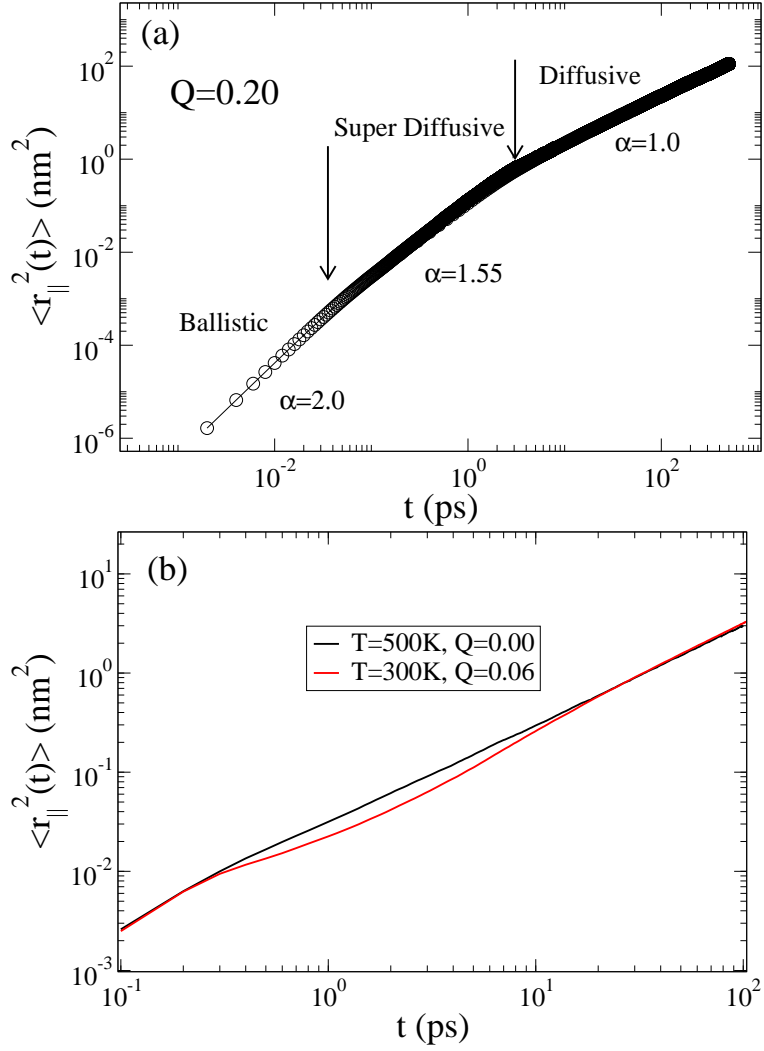

\begin{center}
\includegraphics[width=10cm]{fig4a.eps}
\includegraphics[width=10cm]{fig4b.eps}
\end{center}
\caption{(Color online) (a) Lateral mean square displacement $\langle
r_{||}^2(t)\rangle$ for $Q=0.20$ showing three distinct regimes in the
dynamics -- small time ballistic motion, intermediate time
superdiffusive, and long time diffusive behaviors. (b) Comparison of
MSD at high $T$ and $Q=0$ with MSD for $T=300$ and $Q=0.06$ suggest
that while at high $T$ and $Q=0$ the MSD changes from ballistic to
diffusive behavior as expected for bulk liquids, the MSD shows a
different behavior at the intermediate time scales in the case of
polar confining surfaces.}
\label{msdalltimes}
\end{figure}

\newpage

\newpage

\newpage

\newpage

\end{document}